\documentclass[conference]{IEEEtran}

  	\usepackage[pdftex]{graphicx}
  	\graphicspath{{./pdf/}{./jpeg/}}
	\DeclareGraphicsExtensions{.jpg,.jpeg,.png}

	\usepackage[cmex10]{amsmath}
	\usepackage{mathabx}
	\usepackage{algorithmic}
	\usepackage{array}
	\usepackage{mdwmath}
	\usepackage{mdwtab}
	\usepackage{eqparbox}
	\usepackage{url}
\usepackage{blindtext}
\usepackage{hyperref}
\usepackage[utf8]{inputenc}
\usepackage{booktabs}

\begin{document}

\title{\Large \bf{Initial experiences with Direct Imaging of Neuronal Activity (DIANA) in humans}}


 \author{
 \authorblockN{Shota Hodono$^{1*}$, Reuben Rideaux$^{2,3*}$, Timo van Kerkoerle$^{4}$, Martijn A. Cloos$^{1,5}$}
 \vspace{0.5cm}
 \authorblockA{$^{1}$Centre for Advanced Imaging, The University of Queensland, Australia}
 \authorblockA{$^{2}$Queensland Brain Institute, The University of Queensland, Australia}
 \authorblockA{$^{3}$School of Psychology, The University of Sydney, Australia}
 \authorblockA{$^{4}$Cognitive Neuroimaging Unit, CEA, INSERM, Université Paris-Saclay, NeuroSpin center, 91191 Gif/Yvette, France}
 \authorblockA{$^{5}$ARC Training Centre for Innovation in Biomedical Imaging Technology (CIBIT), The University of Queensland, Australia}
 \authorblockA{$^{*}$SH and RR contributed equally to this work. }
 
 }

\maketitle
\begin{abstract}
Functional MRI (fMRI) has been widely used to study activity patterns in the human brain. It infers neuronal activity from the associated hemodynamic response, which fundamentally limits its spatial and temporal specificity. In mice, the Direct Imaging of Neuronal Activity (DIANA) method revealed MRI signals that correlated with intracellular voltage changes, showing high spatial and temporal specificity. In this work we attempted DIANA in humans. Four visual paradigms were tested, exploring different stimulus types (flickering noise patterns, and naturalistic images) and stimulus durations (50-200ms). Regions of interest (ROI) were derived from traditional fMRI acquisitions and anatomical scans. When using small manually drawn ROI, signals were detected that resembled possible functional activity. However, increasing the stimulus duration did not lead to corroborating signal changes. Moreover, these signals disappeared when averaged over larger functionally or anatomically derived ROI. Further analysis of the data highlighted, DIANA’s sensitivity to inflow effects and subject motion.  Therefore, care should be taken not to mistake artifacts for neuronal activity.  Although we did not yet observe clear DIANA signals in humans, it is possible that a higher spatial resolution is needed to separate signals with opposite signs in different laminar compartments. However, obtaining such data may be particularly challenging because repetitive experiments with short interstimulus intervals can be strenuous for the subjects. To obtain better data, improvements in sequence and stimulus designs are needed to maximize the DIANA signal and minimize confounds. However, without a clear understanding of DIANA’s biophysical underpinnings it is difficult to do so. Therefore, it may be more effective to first study DIANA’s biophysical underpinnings in a controlled, yet biologically revenant, setting.
\end{abstract}

\vspace{0.5cm}

\section{Introduction}
The development of functional magnetic resonance imaging (fMRI) in the 90’s revolutionized neuroscience, offering a way to non-invasively map human brain function \cite{ogawa1990oxygenation,logothetis2008we,bandettini2007functional}. Yet, fMRI’s dependence on changes in the Blood Oxygen Level Dependent (BOLD) signal as a surrogate for neuronal activity limits its spatial and temporal specificity \cite{turner2002much,polimeni2010laminar,zhao2004cortical,kim2003high}.  

In particular, Gradient Recalled Echo (GRE)-BOLD signals are biased toward the larger drainage veins near the pial surface far away from the site of neuronal activity \cite{turner2002much}. Combining Ultra-High field MRI with advanced techniques such as Spin-Echo (SE)-BOLD \cite{zhao2004cortical,han2021improvement,koopmans2019strategies}  or Vascular Space Occupancy (VASO) \cite{huber2017high,yu2019layer} can be used to shift the sensitivity towards the capillary network, closer to the area of neuronal activity. However, even capillary can respond to activity of neurons that are relatively far away \cite{chen2011high}. Still, optical measurements of neuronally driven hemodynamics suggest the spatial specificity of fMRI has not yet reached the physiological limit \cite{drew2011fluctuating,sirotin2009spatiotemporal,hillman2014coupling}.

Perhaps fMRIs biggest limitation is its temporal specificity. Although surprisingly responsive \cite{lewis2016fast,hodono2022tracking}, the hemodynamic response function is sluggish compared to the rapid fluctuations in activity observed at the level of neurons \cite{friston1994analysis}. Invasive techniques in animal models have already indicated that many processes in the brain are so fast and confined to such small areas that they won’t be accessible with BOLD fMRI \cite{steinmetz2019distributed, siegel2015cortical}. Animal studies contribute useful knowledge, however, the ultimate goal is to understand the human brain, where invasive techniques are generally not feasible. Therefore, a non-invasive method that can provide measurements of neuronal activity with high spatiotemporal resolution would be a valuable tool. 

Over the years, various approaches have been proposed to enable more direct observations of neuronal activation through MRI \cite{le2006direct,stanley2018functional, patz2019imaging,roth2023can}. However, each of these methods face challenges of their own. Diffusion weighted fMRI \cite{le2006direct} is easily overshadowed by BOLD effects \cite{miller2007evidence,hodono2022Detection}, and specific absorption rate (SAR) and peripheral nerve stimulation (PNS) considerations make measurements with sub-second temporal resolution difficult. Functional MR spectroscopy \cite{stanley2018functional}, observes activation related variations in metabolite concentrations. This is also subject to experimental considerations such as SAR and signal to noise ratio (SNR), which limit its temporal specificity to $\sim$1s. Elastography based functional MRI \cite{patz2019imaging} provides access to high frequency neuronal activity, but its model-based reconstruction, relying on spatial derivatives of the signal, make it difficult to obtain a high degree of spatial specificity. 

Recently Toi et al. introduced a new MRI based method that aims to enable Direct Imaging of Neuronal Activity (DIANA) with both high temporal and spatial specificity \cite{toi2022vivo}. In their work, they showed signal changes in anesthetized mice that closely follow electrophysiological recordings.

To resolve the mystery of the human mind, non-invasive techniques that provide high spatiotemporal specificity are needed. DIANA may be able to fulfill this role if it can be translated from animals to human experiments \cite{van2022creating}. Here we describe our initial experience attempting to observe neuronal activation in humans using DIANA.

\vspace{0.5cm}

\section{Methods}
\subsection*{Simulations}
The DIANA method is based on a Spoiled Gradient Recalled Echo (SPGRE) sequence. Depending on the exact SPGRE sequence parameters it can take many repetitions for the magnetization to stabilize. Full Bloch simulations were performed using different T1 values to estimate the number of dummy pulses needed to reach the steady state, assuming sequence parameters from the original DIANA paper (TR$=$5ms, flip angle (FA)$=$4$^\circ$ \cite{toi2022vivo}). The MATLAB (MathWorks, USA) code used in these simulations can be found at \href{https://bitbucket.org/shotahodono/diana_spgre_sim}{bitbucket.org/shotahodono}.  

\subsection*{DIANA sequence implementation and setup}
\begin{figure}[h!]

\centering
\includegraphics[width=0.9\columnwidth]{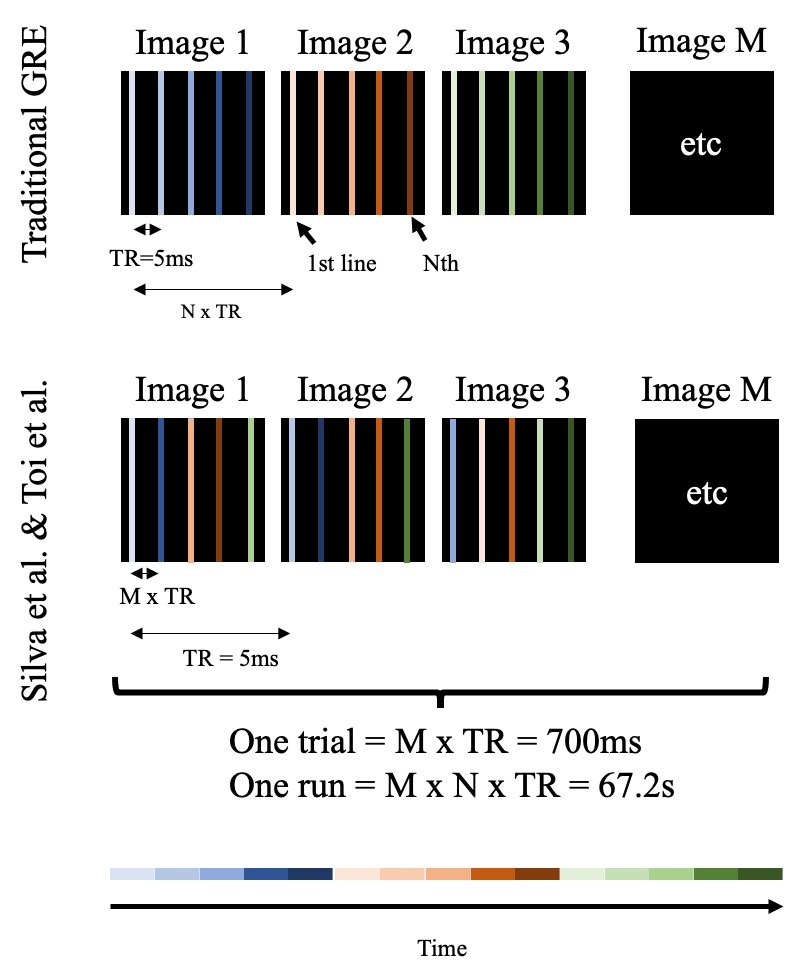} 
\caption{\small Illustrations of traditional SPGRE and DIANA/ Silva-Style
 acquisitions. In a traditional SPGRE acquisition, each second line of k-space in image 1 is acquired immediately after the first. However, in the DIANA approach, the second line is acquired after $M$ TR.}
\label{fig1}
\end{figure}

Our implementation of the DIANA sequence was based on a product SPGRE sequence using 50$^\circ$ quadratic phase increments. As originally proposed by Silva et al. \cite{silva2002laminar}, the DIANA sequence swaps the phase and measurement loops (Fig. \ref{fig1}). Combined with a synchronized repetitive functional paradigm, it becomes possible to obtain extremely high temporal resolutions (e.g. TR$=$5ms), especially when used to image a single slice. Under these conditions, each trial samples the same line in k-space $M$ times once for each image in the final time series. Trials are then repeated $N$ times (number of phase encoding lines), each adding one line to the time series. Thus, collectively it takes at least $M$$\times$$N$$\times$TR to collect one fully sampled dataset, hereafter referred to as “run”.  

To ensure a stable baseline signal, an option was added to enable sufficient dummy pulses to reach the steady state, and trigger signals were added to synchronize the acquisition and functional paradigm. All DIANA experiments were performed with 2$\times$2mm$^2$ in-plane resolution (TR$=$5ms, TE$=$2.4ms, and FA$=$4$^\circ$) at 7 Tesla (Siemens Magnetom, Germany) using a 32-channel head coil (Nova Medical, USA).

\subsection*{Phantom experiments}
Phantom experiments were performed to evaluate the stability of the MRI signal. After allowing fluid motion to settle ($\sim$30 min), a custom phantom containing 50mL centrifuge tubes with different concentrations of manganese chloride was imaged using both the SPGRE and DIANA sequences. The protocol was chosen such that both measurements produced an equal number of readouts. The SPGRE collected 1024 sequential images in two runs, the first 21 measurements ($\sim$2000 TR) were removed to ensure that the SPGRE signal reached the steady state. Two scans of 4 runs of DIANA measurements were collected with 700ms trials, using 2000 dummy TRs to stabilize the signal at the start.    

\subsection*{DIANA paradigms}
Four different paradigms were tested, each in a different scan session (summarized in Table \ref{table:1}). Data for paradigms I-III was collected using a single oblique axial slice centred on the calcarine sulcus (Fig. \ref{fig2}). Data for paradigm IV was collected using a single oblique sagittal slice offset from the great longitudinal fissure (Fig. \ref{fig2}).

Each trial in the DIANA paradigm consisted of a 50-200ms visual stimulus with a 500-600ms interstimulus interval (ISI) (Table \ref{table:1}). Visual stimuli were either pseudo-randomly configured noise patterns or naturalistic images that changed configuration on each trial. During ISIs, a blank (black/grey) screen was presented to minimize visual stimulation, e.g., caused by blinking. Lights in the scanner room were dimmed to reduce contrast between ambient light and closed eyes. In total, 11 runs were collected (1056 trials, 1 scan of 11 runs, 15min of DIANA scan time per subject) for paradigms I and II, and 33 runs were collected (3168 trials, 3 scans of 11 runs, 45min of DIANA scan time per subject) for paradigms III and IV. 
\begin{table}[h!]
\caption{Paradigm configurations and numbers of subjects.}
\begin{tabular}{c|c|c|c|c}
    & Type & on [ms] & ISI [ms]  &$\#$ of subjects \\
  \bottomrule[1pt]
Paradigm I& noise& 50 & 550 & 1\\
 \hline
Paradigm II& noise& 200 & 500 & 1\\
 \hline
Paradigm III& noise& 100 & 600 & 3\\
 \hline
Paradigm IV& nat. img.& 100 & 600 & 1\\
 \hline
\end{tabular}
\label{table:1}
\end{table}

Three human adult males and one female (23-40yo) participated in the experiments, having provided written informed consent. All participants had either normal or corrected-to-normal vision and were screened for MRI contraindications prior to scanning. The study was approved by the local human research ethics committee in accordance with national guidelines.

\subsection*{In-vivo slice placement}
Our objective was to observe clear and reproducible neuronal signals using DIANA. Considering that Toi et al reported $\sim$0.1\% signal change \cite{toi2022vivo}, we decided to avoid parallel imaging strategies that could introduce subtle artifacts \cite{robson2008comprehensive}. Instead, we focused our attention on a single slice and maximized our sensitivity through averaging. 

\begin{figure}[h!]
\centering
\includegraphics[width=0.9\columnwidth]{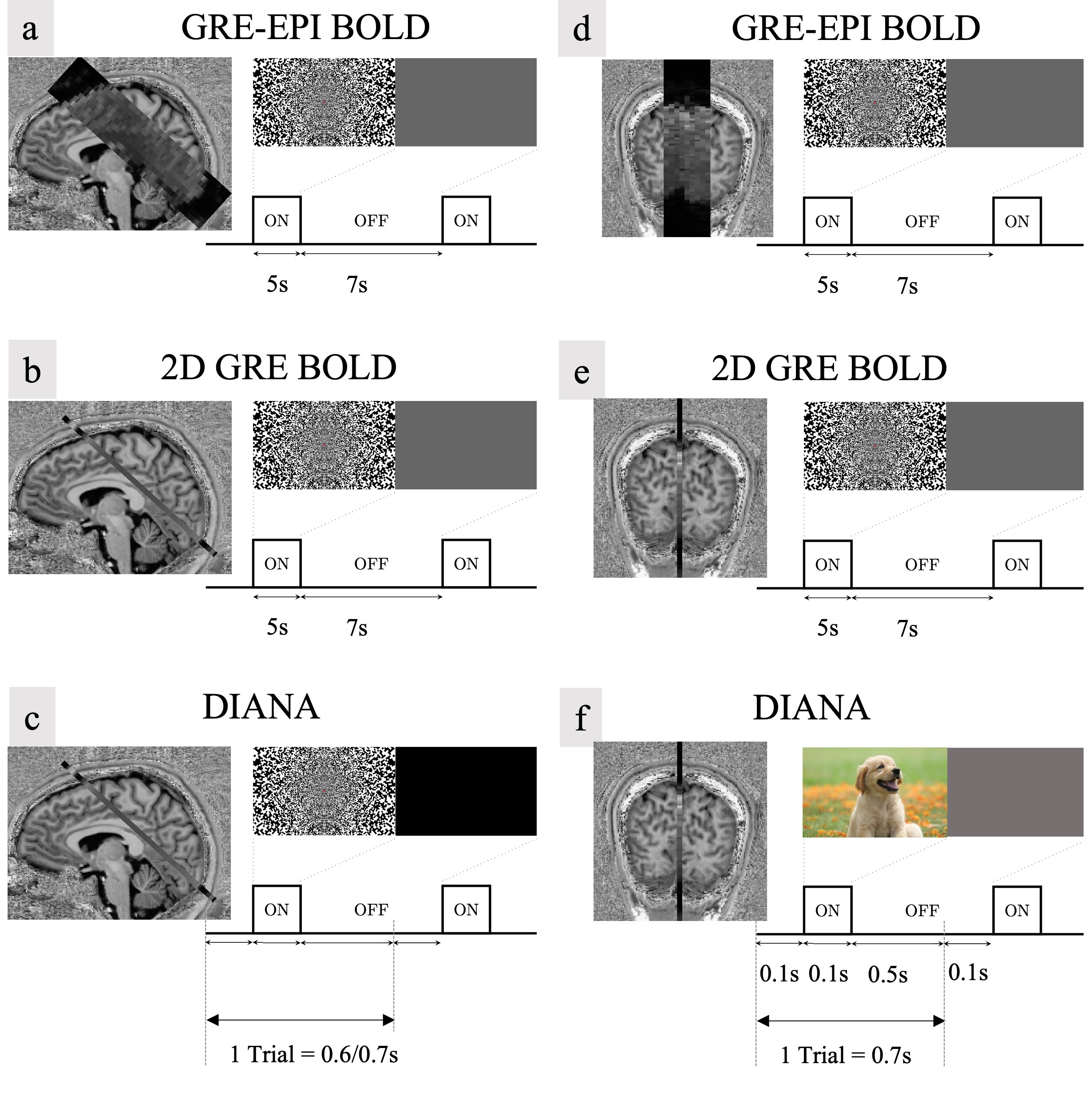} 
\caption{\small Slice placement and paradigm designs for in vivo experiments. (a) GRE-EPI BOLD for slice placement in paradigm I, II, and III. (b) GRE BOLD for functional localization in paradigm III. (c) DIANA acquisition in paradigm I, II, and III. (d) GRE-EPI BOLD for slice placement in paradigm IV. (b) GRE BOLD for functional localization in paradigm IV. (c) DIANA acquisition in paradigm IV. }
\label{fig2}
\end{figure}

The DIANA imaging slice was identified based on a BOLD- based GRE-EPI functional localizer (9 slices, TR$=$1s, TE$=$20ms, FA$=$60$^\circ$, 2$\times$2$\times$5mm, 360 volumes). A visual paradigm was used (5s on 7s off, Fig. \ref{fig2}a) and quickly analyzed using the Fourier-transform (along the measurement dimension) to identify voxels that match the expected spectral function based on the convolution of the canonical hemodynamic response function with the functional paradigm \cite{hodono2022tracking}.

\subsection*{Functional localization}
In paradigms III and IV, the target slice was also imaged using a single slice SPGRE sequence (TR$=$31ms, TE$=$20ms, FA$=$10$^\circ$, 2$\times$2$\times$5mm, GRAPPA$=$3, 360 volumes) to obtain a BOLD based map of functional activity without geometric distortion (5s on 7s off, Fig. \ref{fig2}b). The z-score maps were obtained through general linear modelling analysis with FSL (https://fsl.fmrib.ox.ac.uk/fsl/). Voxels with z-scores above 5 (in the general vicinity of V1), were used to identify BOLD based regions of interest (ROI) where a DIANA signal may be expected.

Given the relatively low spatial resolution used in this study (2$\times$2$\times$5mm), reasonable voxel-wise coincidence between BOLD and DIANA signals may be expected, even though the GRE BOLD signal is weighted towards the drainage veins at the surface \cite{turner2002much,polimeni2010laminar}. Nevertheless, we also collected a T1 weighted image of the target slice using a 2D adaptation of the MP2RAGE sequence \cite{marques2010mp2rage}. These images were used to manually draw anatomically informed ROI using ITK-SNAP \cite{yushkevich2006user}.

\subsection*{Image reconstruction \& Data analysis}
All DIANA data were reconstructed offline with 16bit dynamic range (MATLAB, MathWorks, USA). The raw signal measured in each voxel for every trial was first converted to a percent signal change as a function of time, then detrended (linear), and smoothed with a gaussian kernel (width $=$ 3 time points). Percent signal change was then averaged across ROI.

\subsection*{Evaluation of physiological noise in DIANA}
To investigate physiological noise contribution to the DIANA acquisition, one of the four subjects was invited to a further scanning session. The subject was instructed remain awake during the scan, but no visual stimuli were presented during this session. The same experimental parameters were used as in Paradigm III, but slice thickness was varied from 8mm to 2mm. The data were reconstructed in the same manner as the functional DIANA data. The data were then Fourier transformed for further artifact analysis.

\vspace{0.5cm}
\section{Results and Discussion}
\subsection*{Simulations}
\begin{figure}[h!]
\centering
\includegraphics[width=0.9\columnwidth]{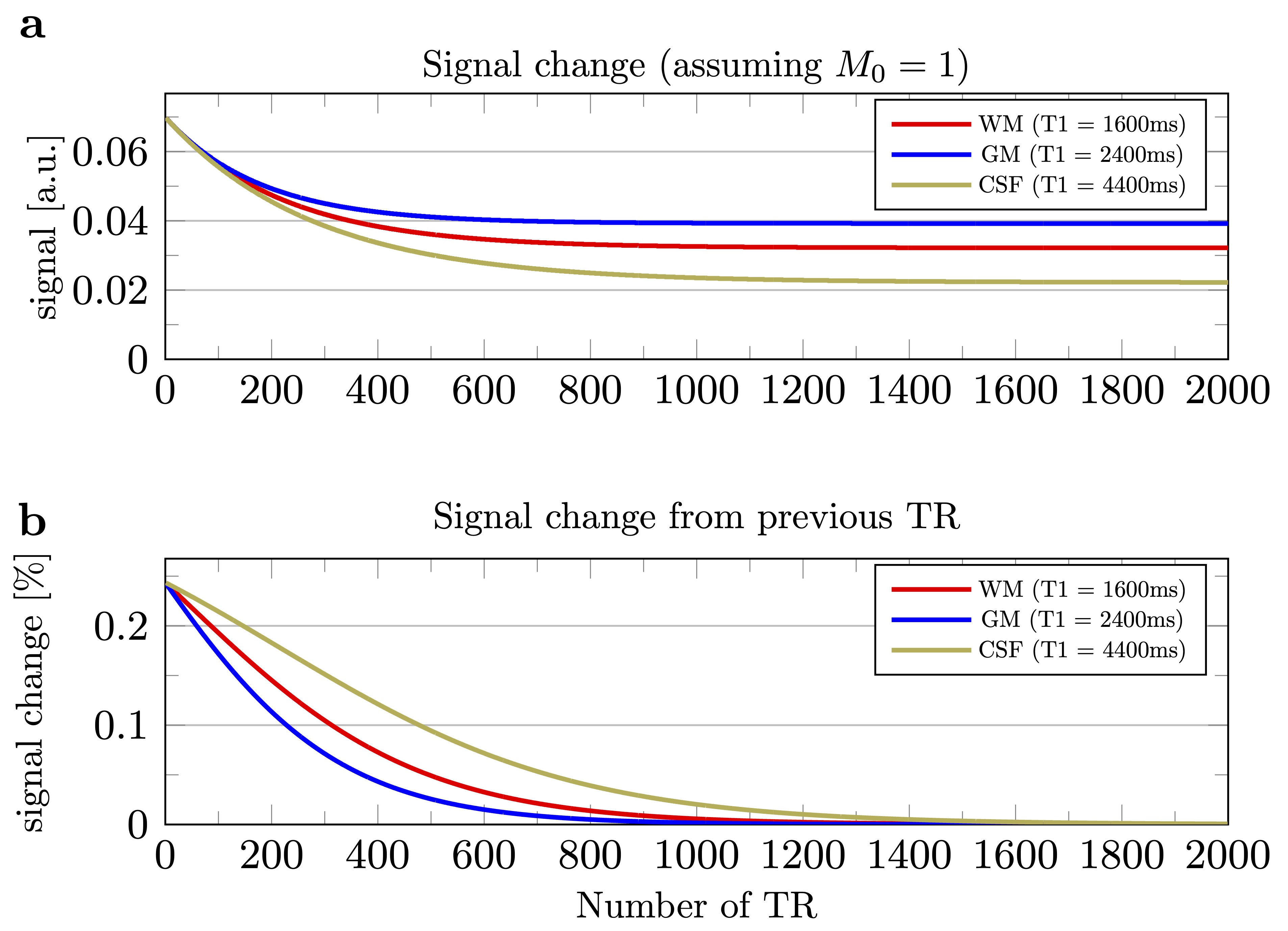} 
\caption{\small Simulations results showing the convergence towards the steady state signal. The simulated SPGRE signal (assuming unit equilibrium magnetization) as a function of pulse repetition (a). The change in SPGRE signal as a function of pulse repetition (b).}
\label{fig3}
\end{figure}

Bloch simulations indicated that 1000 to 1500 dummy pulses were needed to reach the steady state for all tissues (Fig. \ref{fig3}). With a 5ms TR, this translates to 10s worth of dummy pulses. Given that it takes about 67s to complete a single DIANA experiment with a 96$\times$96 matrix and 140 time points ($N$$=$96, $M$$=$140), it is considerably more efficient to complete multiple DIANA measurements per scan. In our experiments, we ran 11 runs per scan ($\sim$12.5min).

\subsection*{Phantom experiments}
\begin{figure}[h!]
\centering
\includegraphics[width=0.9\columnwidth]{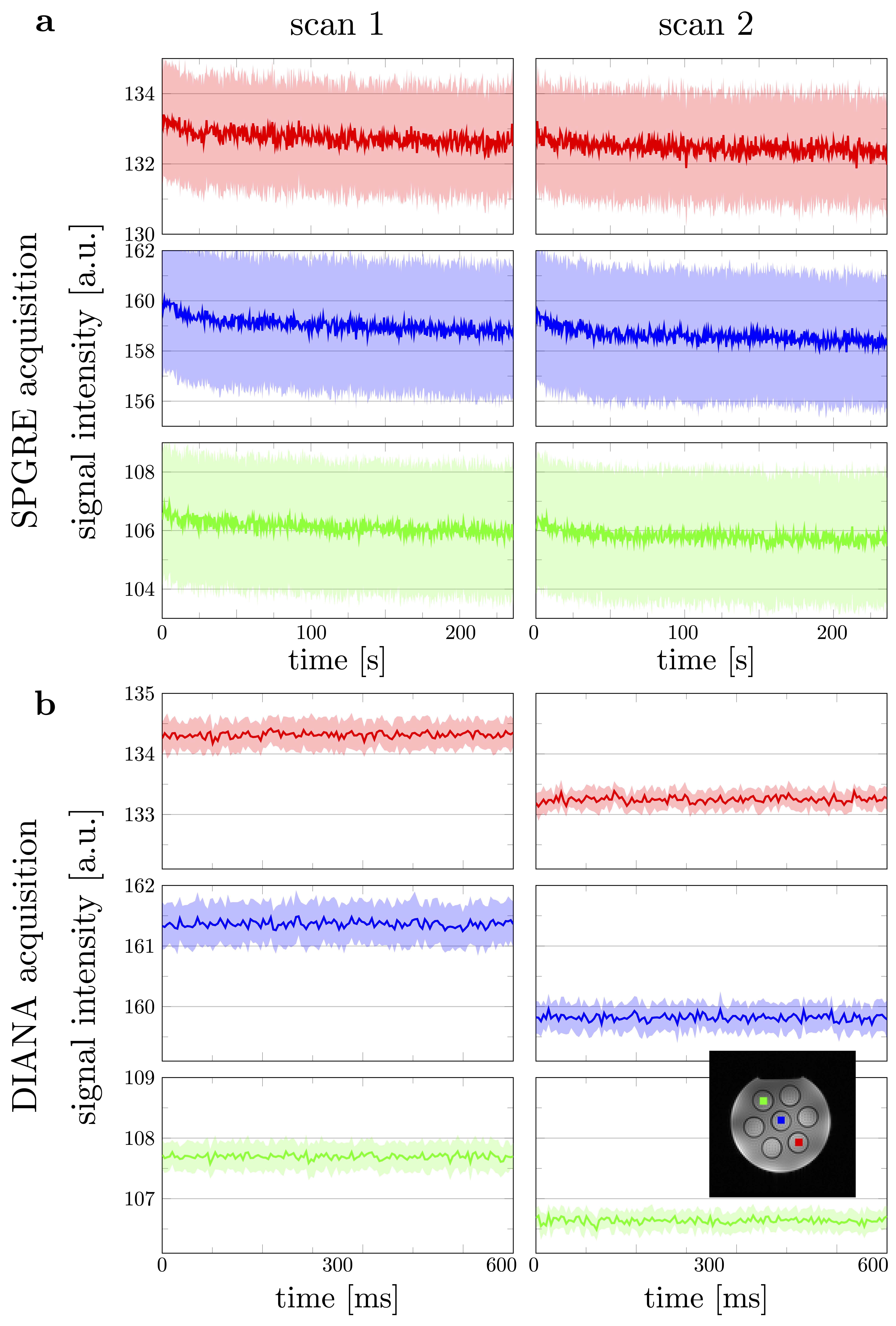} 
\caption{\small Phantom measurements showing the effect of scanner drift on traditional SPGRE and DIANA sequences. Two sequentially collected traditional SPGRE measurements, showing a gradual drift in signal (a). Two sequentially collected DIANA measurements, showing a step change in signal (b). Both measurements comprise equal number of readouts. Highlighted areas around the curve indicate the 95\% confidence interval across voxels for (a) and across runs and voxels for (b). }
\label{fig4}
\end{figure}

The mean signal in each image is dominated by the center for k-space. The standard SPGRE sequence passes through the center of k-space once every $N\times$TR, such that subsequent measurements directly reflect scanner drift (Fig. \ref{fig4}a). Interestingly, although simulations suggested that 2000 dummy TR were adequate to reach the steady state, some samples still showed signs of a residual transient. Therefore, in the following DIANA analysis, we decided to discard the first run from each scan, in addition to the initial 2000 dummy TRs.

The DIANA sequence rapidly passes through the center of k-space for all time points in one trial, but then will not revisit the k-space center until $N\times M \times$TR later. Consequently, signal drift now presents itself as a step function every $N\times M \times$TR (Fig. \ref{fig4}b). Importantly, because each trial is individually normalized based on its mean signal before averaging, these drifts are effectively eliminated from the DIANA analysis.  Confidence intervals for SPGRE acquisition were calculated across voxels within ROI. Whereas, in the DIANA acquisition, mean signal over the ROI was first calculated and the confidence intervals were calculated across runs. The different procedures resulted in wider confidence intervals in SPGRE acquisition.

\subsection*{Paradigms I \& II}
\begin{figure}[h!]
\centering
\includegraphics[width=1 \columnwidth]{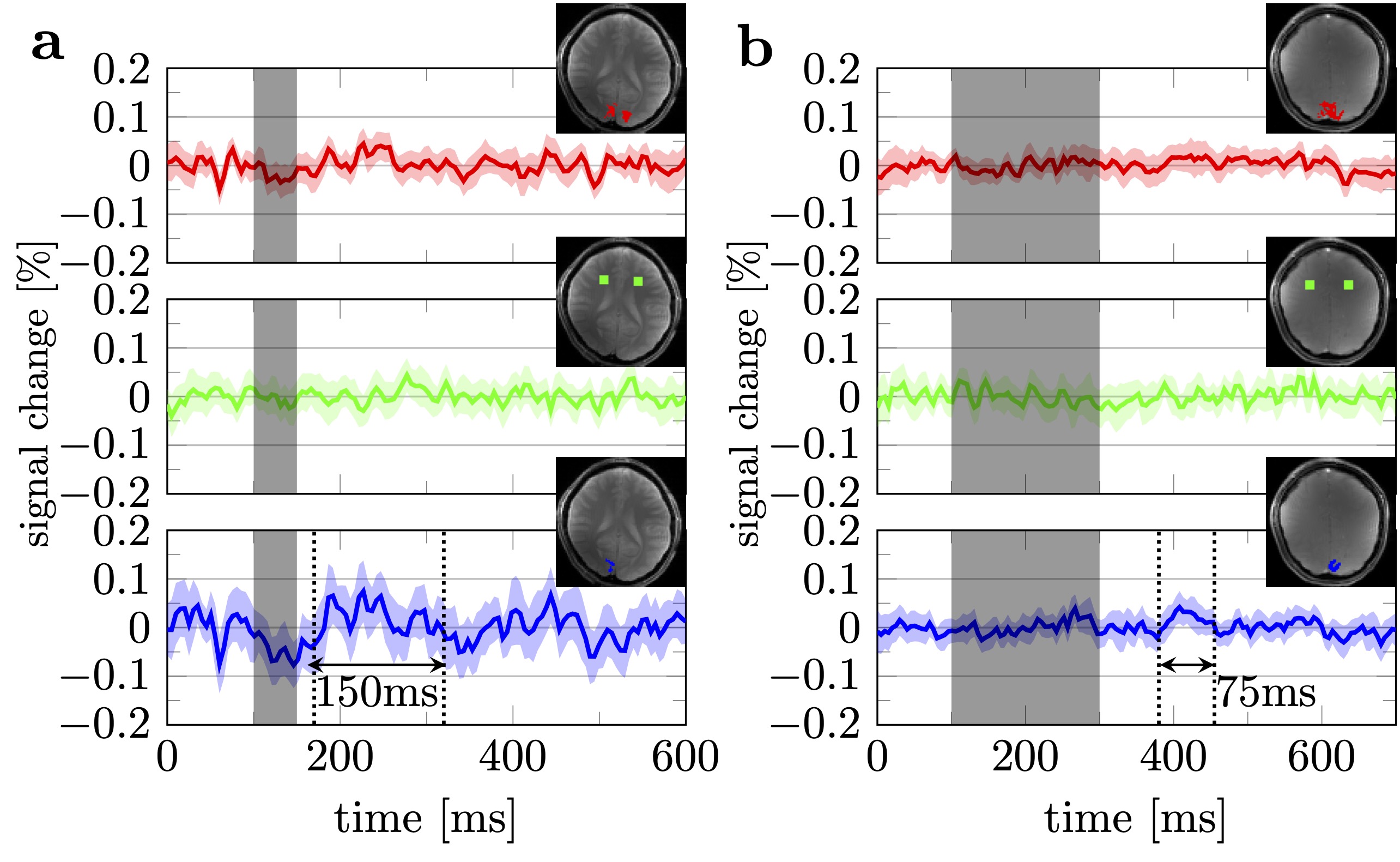} 
\caption{\small DIANA results using paradigm I \&II.}
\label{fig5}
\end{figure}

The DIANA signal obtained with Paradigm I showed some promise in a local ROI (Fig. \ref{fig5}a). A 0.05\% increase in signal, approximately half of that was observed in mice \cite{toi2022vivo}, followed the stimulus onset by $\sim$75ms. This signal increase persisted for $\sim$150ms.

To test whether the observed signal was related to the stimulus, we increased the stimulus duration from 50 to 200ms. Again, some localized areas showed a peak in the time averaged signal (0.05\%) after the stimulus (Fig. \ref{fig5}b). However, while the stimulus duration was longer, the duration of the putative signal peak was reduced. By contrast, if the change in activity was related to the stimulus, one would expect to observe a prolonged peak exceeding the duration of that observed in Paradigm I (150ms) \cite{mirpour2009state}. It should also be noted that the onset of the putative signal peak in Paradigm II is inconsistent with the temporal dynamics of the visual processing cascade. That is, the peak occurs $\sim$300ms after the stimulus onset, whereas responses in early visual cortex are evoked between 50 to 100ms after the stimulus onset \cite{ringach2003dynamics}. 

\subsection*{Paradigms III}

\begin{figure}[h!]
\centering
\includegraphics[width=0.9\columnwidth]{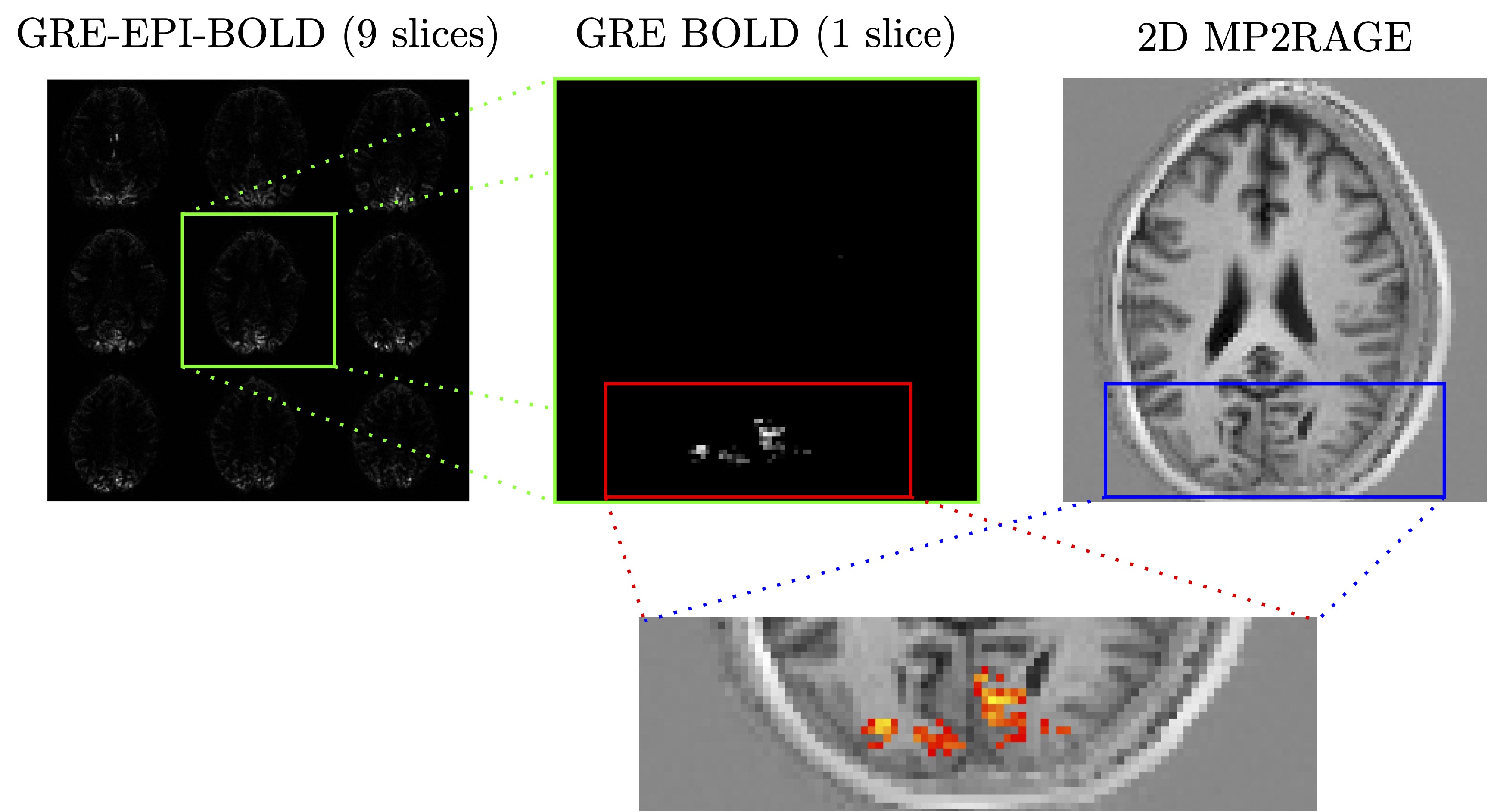} 
\caption{\small Exemplary functional and anatomical localization data from an oblique axial experiment. The top left panel shows the activation estimate based on the Fourier analysis of GRE-EPI BOLD images. The middle panel shows a z-score map obtained using single slice SPGRE-BOLD. The right panel shows the T1 weighted anatomical matching the target slice. The bottom panel shows the z-score ($>$5) derived activation superimposed on the 2D adaptation of the MP2RAGE. }
\label{fig6}
\end{figure}
\begin{figure}[h!]
\centering
\includegraphics[width=0.8\columnwidth]{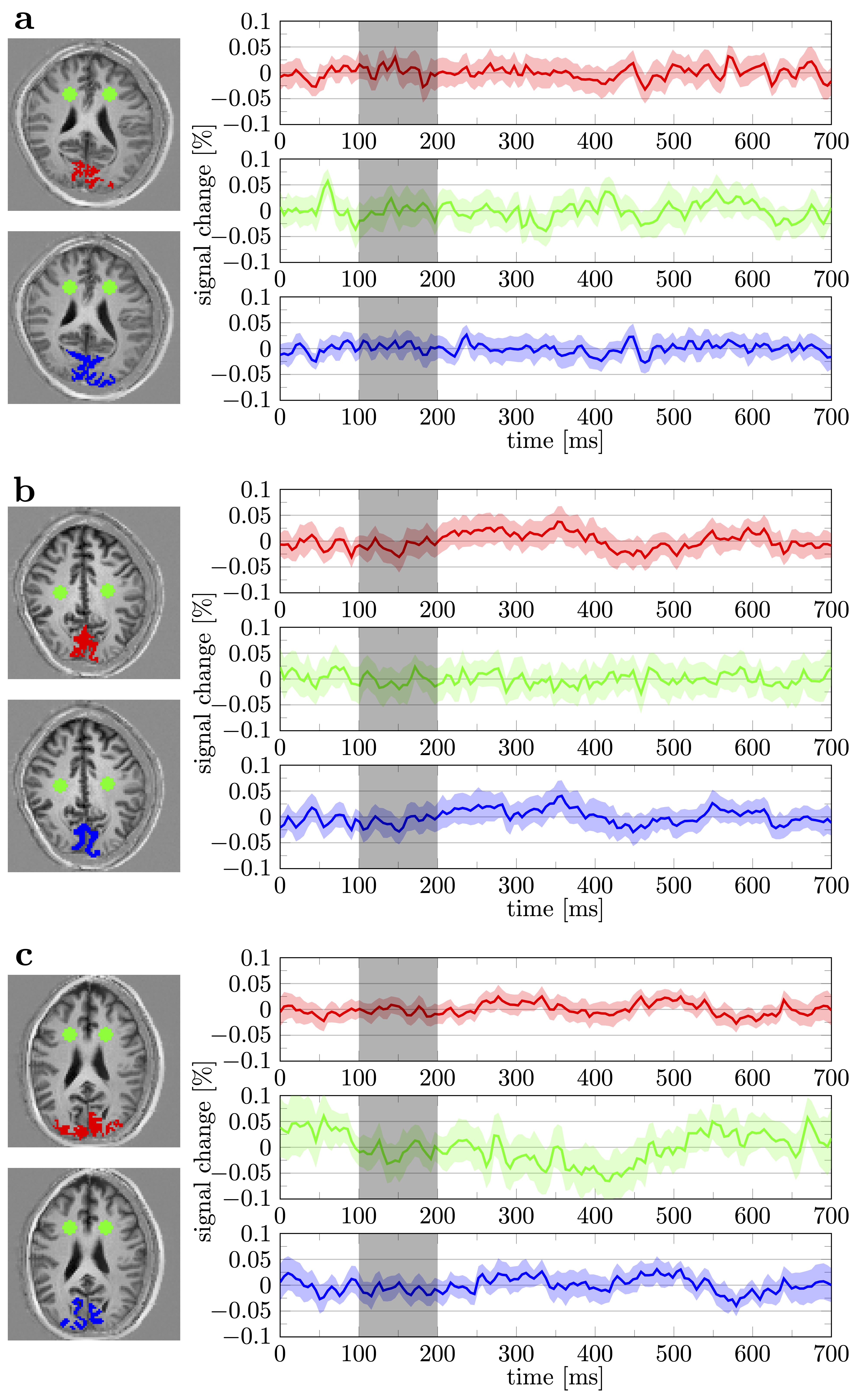} 
\caption{\small BOLD based functional localizers and DIANA results using paradigm III. The red curves show the trial averaged DIANA signal obtained using the BOLD ROI. The blue curves show the trial averaged DIANA signal obtained using hand drawn ROI following the anatomy. Shaded areas show the 95\% confidence interval.}
\label{fig7}
\end{figure}
The results of paradigms I and II suggest spurious results may be found, possibly due to limited signal averaging power and poorly defined ROI. So, we sought to collect more data, and repeat the same paradigm 3 times, across different participants. Further, in the following set of experiments we also collected distortion matched SPGRE-BOLD and T1-weighted anatomical data to improve the ROI definition.

Each participant showed clear signs of stimulus related activation across the visual cortex in the EPI based functional localizer (Fig. \ref{fig6}, left). In all cases, the target slice also showed clear activation in the single slice distortion free 2D GRE data (Fig. \ref{fig6}, middle), matching the expected anatomy.

When averaged across BOLD (Fig. \ref{fig7} Red) or anatomical ROI (Fig. \ref{fig7} Blue), trial averaged DIANA data showed no signs of activation. Moreover, the local ROI (based on both hand-drawn and BOLD) failed to show plausible signs of stimulus related activation. 

Volunteers reported that the noise-like checkerboard stimuli with ISI $\sim$500ms were intense, bordering uncomfortable, motivating them to blink in anticipation. In general, blinking may be a confound in these visual paradigms using rapid stimulus presentation. A typical eye blink lasts $\sim$100ms, creating a visual contrast change in similar duration to the target stimulus. Although blinking could be detected using an eye tracker, data scrubbing may be difficult because DIANA inherently averages trials during image reconstruction.

\subsection*{Paradigms IV}
Based on our experiences in Paradigms I-III we surmised that tailored naturalistic images, which are more engaging and comfortable to view, may be a superior visual stimulus for the experiment. In addition, we reasoned that a strategically placed sagittal slice would reduce partial volume effects.

\begin{figure}[h!]
\centering
\includegraphics[width=0.9\columnwidth]{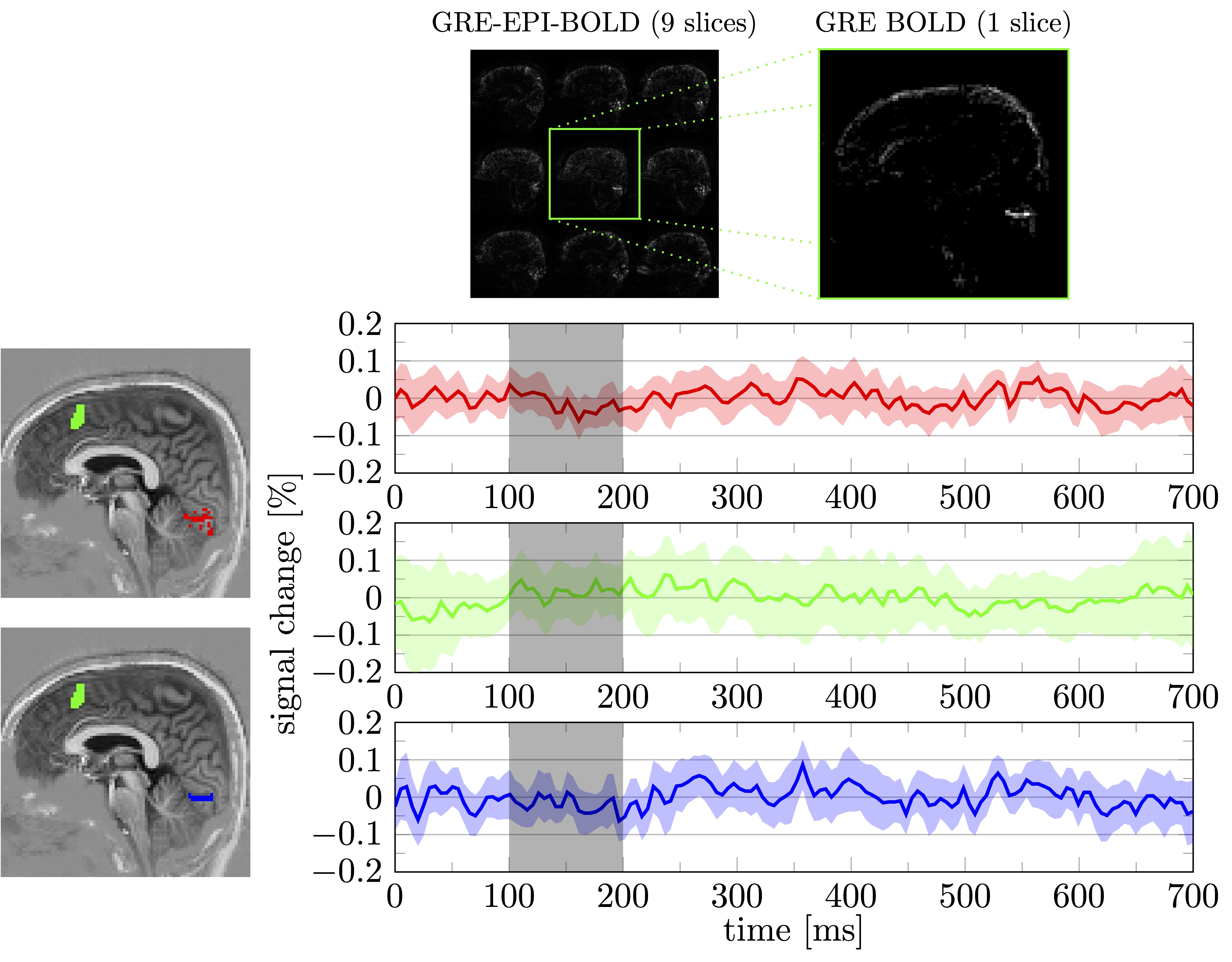} 
\caption{\small BOLD based functional localizers and DIANA results from paradigm IV. The red, green and blue traces show the mean trial averaged DIANA signal within the BOLD, control, and anatomical ROIs. The blue trace shows the trial averaged DIANA signal obtained using hand drawn ROI following the anatomy. Shaded areas show the 95\% confidence interval across runs.}
\label{fig8}
\end{figure}

The participant indeed reported that the naturalistic images were more comfortable, making it easier to remain attentive. Incidentally, incorporating natural images makes it easier to monitor attention. Nevertheless, no convincing evidence of a DIANA signal was observed (Fig. \ref{fig8}).

\subsection*{Notes on physiological noise}
\begin{figure}[h!]
\centering
\includegraphics[width=0.9\columnwidth]{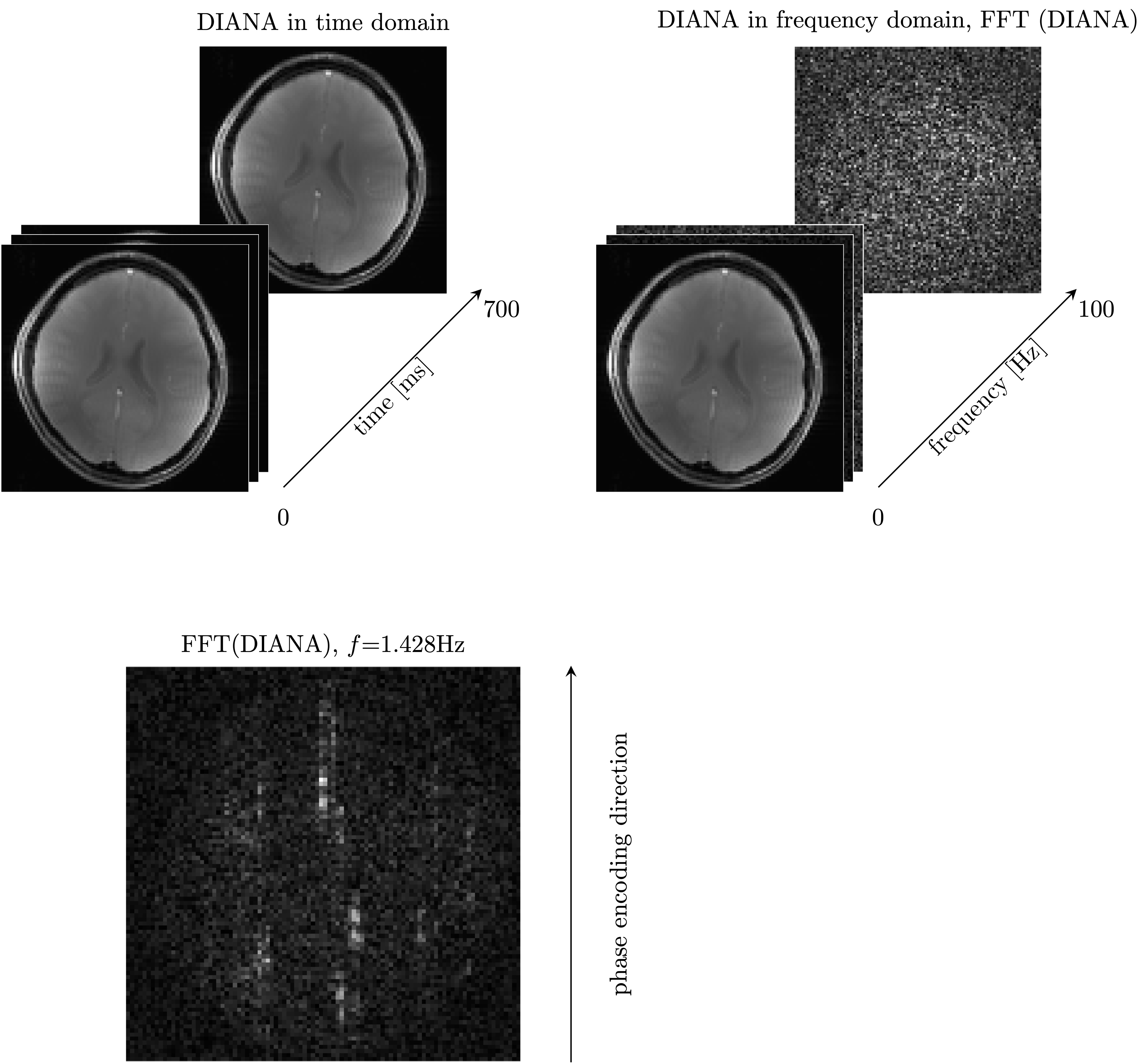} 
\caption{\small Noise analysis in DIANA acquisition. A subject was scanned using same experimental parameters used in paradigm III, only this time no visual stimuli were presented. The acquired images were Fourier transformed to investigate the noise and artifact distribution.}
\label{fig9}
\end{figure}

The magnetization in the imaging slice is strongly saturated. Consequently, inflow effects produce strong blood signals. Taking the Fourier transform of the timeseries, clear flow artifacts can be seen aliased throughout the phase-encoding direction (Fig. \ref{fig9}). Following similar mechanics, cerebrospinal fluid (CSF) pulsation is also likely to produce temporally varying signal intensities. 

\begin{figure}[h!]
\centering
\includegraphics[width=0.9\columnwidth]{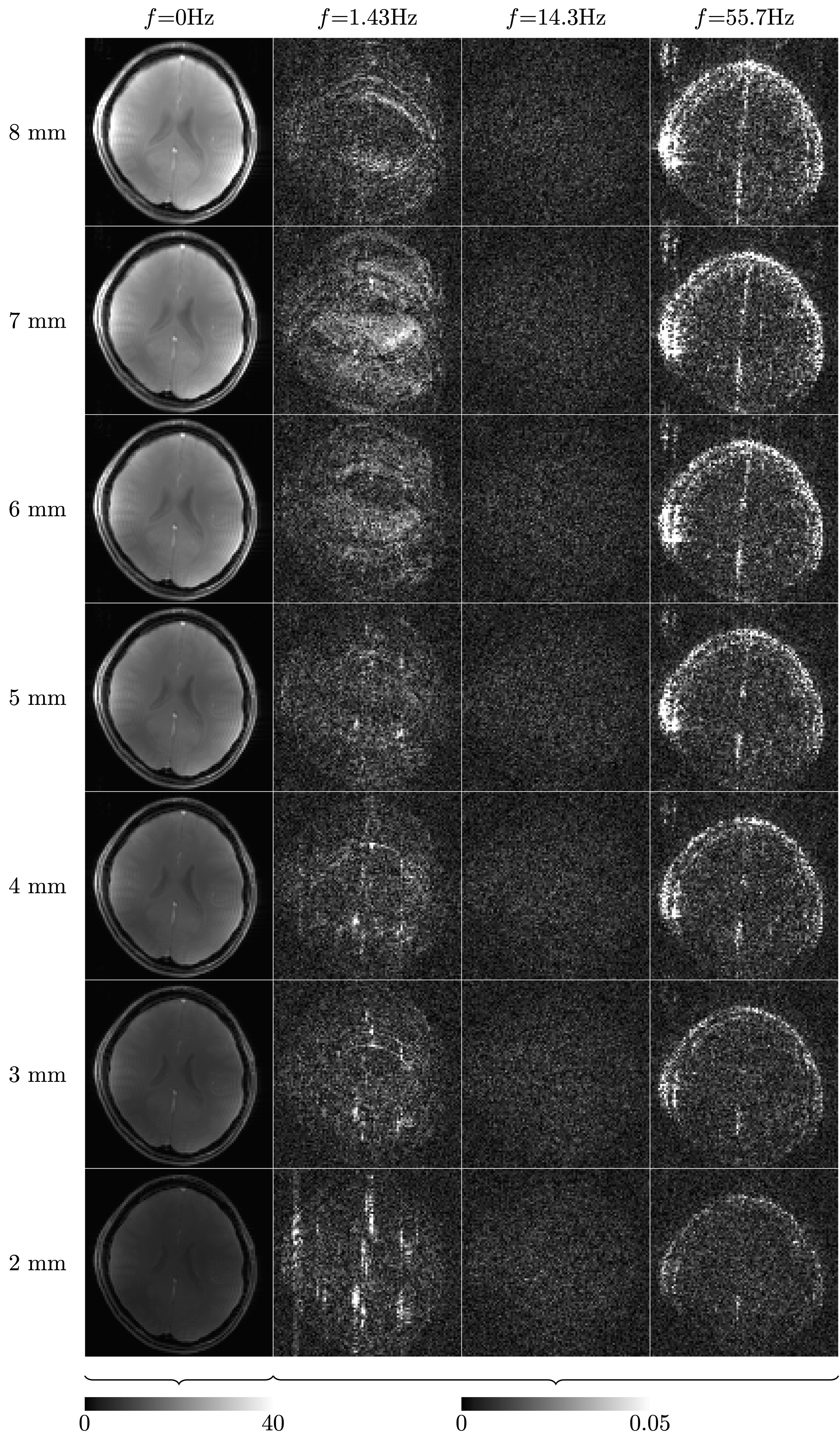} 
\caption{\small DIANA images in the frequency domain, when imaged with different slice thicknesses. }
\label{fig10}
\end{figure}

The frequency analysis also revealed that artifacts are seen at specific frequencies, $f$=1.43Hz and $f$=55.7Hz (Fig. \ref{fig10}). All other frequencies except $f$=0Hz show similar noise patterns as those seen at $f$=14.3Hz. At $f$=55.7Hz, the artifacts are reduced when a thinner slice is used, which suggests that some physiological noise components can be suppressed using a smaller voxel size. However, when using thinner slices vascular inflow effects became more pronounced. The signal intensity of these artifacts was three orders of magnitude smaller than the mean signal ($f$=0Hz), which overlaps with the expected range of DIANA signals ($\sim$0.1\%). Therefore, care must be taken to exclude such spurious signals. An optimized imaging strategy may be required eliminate flow related artifacts.

\subsection*{Recommendations for future experiments}
Clearly, part of the challenge is finding a DIANA paradigm that works well with human subjects. In addition to visual stimuli, we also explored auditory stimuli, which were more comfortable for the subject and could benefit from the high temporal specificity of the auditory system \cite{ zanker2002temporal, gazzaniga2000new}. Nevertheless, auditory stimuli also failed to produce a reliable DIANA signal (data not shown). However, in the auditory experiment, the BOLD response was also notably weaker, presumably due to the auditory noise produced by the scanner itself, which may be even more problematic when using short ISIs. Finger tapping may be an effective alternative. However, care should be taken not to inadvertently induce head motion.   

One other challenge is resolution. In mice, Toi et al \cite{toi2022vivo} found signals confined to different laminar compartments, with a negative transient in superficial layers and positive changes in the middle and deep layers. It might be that a spatial resolution of ~0.5mm is required to resolve such spatially confined signals in humans, somewhat similar to the high resolution that is required to record electro physiological signals from single units \cite{hubel1957tungsten}.  However, obtaining such resolutions with adequate SNR to resolve signal changes of 0.1\% may be challenging.

In addition, our current understanding of DIANA’s biophysical underpinning is still very limited. Without such knowledge it is difficult to make informed decision regarding sequence parameterization and paradigm design. 

To answer such crucial questions, dedicated studies in highly controlled, yet biologically representative, methods may be needed. Revealing the biophysical underpinning of the DIANA signal and validating theories regarding interference between excitatory and inhibitory signals will be a crucial step toward widespread adoption and potential translation to humans.

\vspace{0.5cm}


\section{Conclusion}
The translation of DIANA from animals to humans appears to be non-trivial and care should be taken not to mistake spurious signals for neuronal activity. Nevertheless, considering the potential payoff, continued effort translating DIANA from mice to humans may well be worth it. To aid in this endeavor, parallel studies focused on DIANA’s biophysical underpinning and ability to detect excitatory and inhibitory signals in controlled settings will likely be of great value.

\vspace{0.5cm}


\section{Acknowledgments}
This work was supported by young investigator award from the French National Research Agency (ANR-19-CE37-003-01), the Australian government through Australian Research Council (ARC) Future fellowship grant FT200100329, ARC Discovery Early Career Researcher Award (DE210100790), and ARC Centre grant (IC170100035). The authors acknowledge the facilities of the National Imaging Facility at the Centre for Advanced Imaging. 

\vspace{0.5cm}


\bibliographystyle{IEEEtran}
\bibliography{library}

\end{document}